\documentclass[12pt,onecolumn]{article}	
\usepackage[top=1in, bottom=1in, left=1in, right=1in]{geometry}
\usepackage{graphicx}
\usepackage{amssymb}
\usepackage{amsmath}  				
\usepackage{epstopdf}  				
\usepackage{graphicx,float,wrapfig} 	
\usepackage{color}					
\usepackage[colorlinks,linktocpage=true]{hyperref}		
\usepackage[font=small,labelfont=bf,textfont=sf,
labelsep=quad]{caption}				

\definecolor{purple}{RGB}{152, 66, 227}
\definecolor{magenta}{RGB}{205, 16,118}
\definecolor{turquoise}{RGB}{53, 173, 153}

\usepackage[superscript,space]{cite}	

\hypersetup{						
  citecolor= black,
  linkcolor=black,
}

\usepackage{setspace}   

\usepackage [english]{babel}
\usepackage [english = american]{csquotes}
\MakeOuterQuote{"}

\doublespace                   				

 \usepackage[protrusion=false,expansion=false]{microtype}

\renewcommand{\v}[1]{\ensuremath{\mathbf{#1}}} 
 
\newcommand{\abs}[1]{\left| #1 \right|} 
\renewcommand{\d}[2]{\frac{d #1}{d #2}} 
 
\let\baraccent=\= 
\renewcommand{\=}[1]{\stackrel{#1}{=}} 

\begin{document}

\title{Rapid behavioral transitions produce chaotic mixing by a planktonic microswimmer}
\author
{William Gilpin$^{1\ast}$, Vivek N. Prakash$^{2}$, Manu Prakash$^{2\ast\ast}$\\
\\
\normalsize{$^{1}$Department of Applied Physics, $^{2}$Department of Bioengineering, }\\
\normalsize{Stanford University, Stanford, CA}\\
\\
\normalsize{$^\ast$wgilpin@stanford.edu}, \normalsize{$^{\ast\ast}$manup@stanford.edu}
\\
}

\maketitle

\newpage
\begin{abstract}
Despite their vast morphological diversity, many invertebrates have similar larval forms characterized by ciliary bands, innervated arrays of beating cilia that facilitate swimming and feeding. Hydrodynamics suggests that these bands should tightly constrain the behavioral strategies available to the larvae; however, their apparent ubiquity suggests that these bands also confer substantial adaptive advantages. Here, we use hydrodynamic techniques to investigate "blinking," an unusual behavioral phenomenon observed in many invertebrate larvae in which ciliary bands across the body rapidly change beating direction and produce transient rearrangement of the local flow field. Using a general theoretical model combined with quantitative experiments on starfish larvae, we find that the natural rhythm of larval blinking is hydrodynamically optimal for inducing strong mixing of the local fluid environment due to transient streamline crossing, thereby maximizing the larvae's overall feeding rate. Our results are consistent with previous hypotheses that filter feeding organisms may use chaotic mixing dynamics to overcome circulation constraints in viscous environments, and it suggests physical underpinnings for complex neurally-driven behaviors in early-divergent animals.
\end{abstract}
\newpage


Early animals evolved in the oceans, where strong fluidic constraints guided the evolution of novel morphologies and behavioral strategies\cite{costello2008medusan,barnes2009invertebrates}. Such physical forces partly explain the surprising ubiquity of ciliary bands, linear arrays of beating cilia (controlled by a central nervous system) that facilitate swimming and generate feeding currents in the larval forms of most marine invertebrates\cite{amemiya2015development,gilpin2017vortex}. Given recent intense interest in unravelling the physics of low Reynolds number swimming in unicellular organisms\cite{guasto2012fluid,goldstein2016batchelor}, ciliary bands provide a starting point for studying how a microscale multicellular organism's repertoire of neurally-driven behavior is limited by strong physical constraints.

A behavioral phenomenon observed in the ciliary bands of many invertebrate larvae is "blinking," in which cilia across the body spontaneously and transiently halt or reverse beat direction---causing dramatic but short-lived changes in the swimming stroke readily visible in timelapses of the organism\cite{mackie1976nervous,mackie1969electrical,akira1989control}. In certain organisms these blinks have been shown to play a direct role in regulating swimming depth in the water column\cite{conzelmann2011neuropeptides,mackie1969electrical}. However, some authors have hypothesized that these short-term reversals also play a role in feeding\cite{barnes2009invertebrates,riisgaard2010particle,akira1989control,arkett1987neuronal,pernet2017larval}, a hypothesis consistent with their relatively short duration, regular cadence, and dramatic rearrangement of the local feeding current. Here, we use quantitative behavioral analysis and hydrodynamic modeling to suggest that these short-term blinks enhance larval feeding rates, by mixing the local flow field and thus improving capture efficiency of passing food particles.

\begin{figure}
{
\centering
\includegraphics[width=\linewidth]{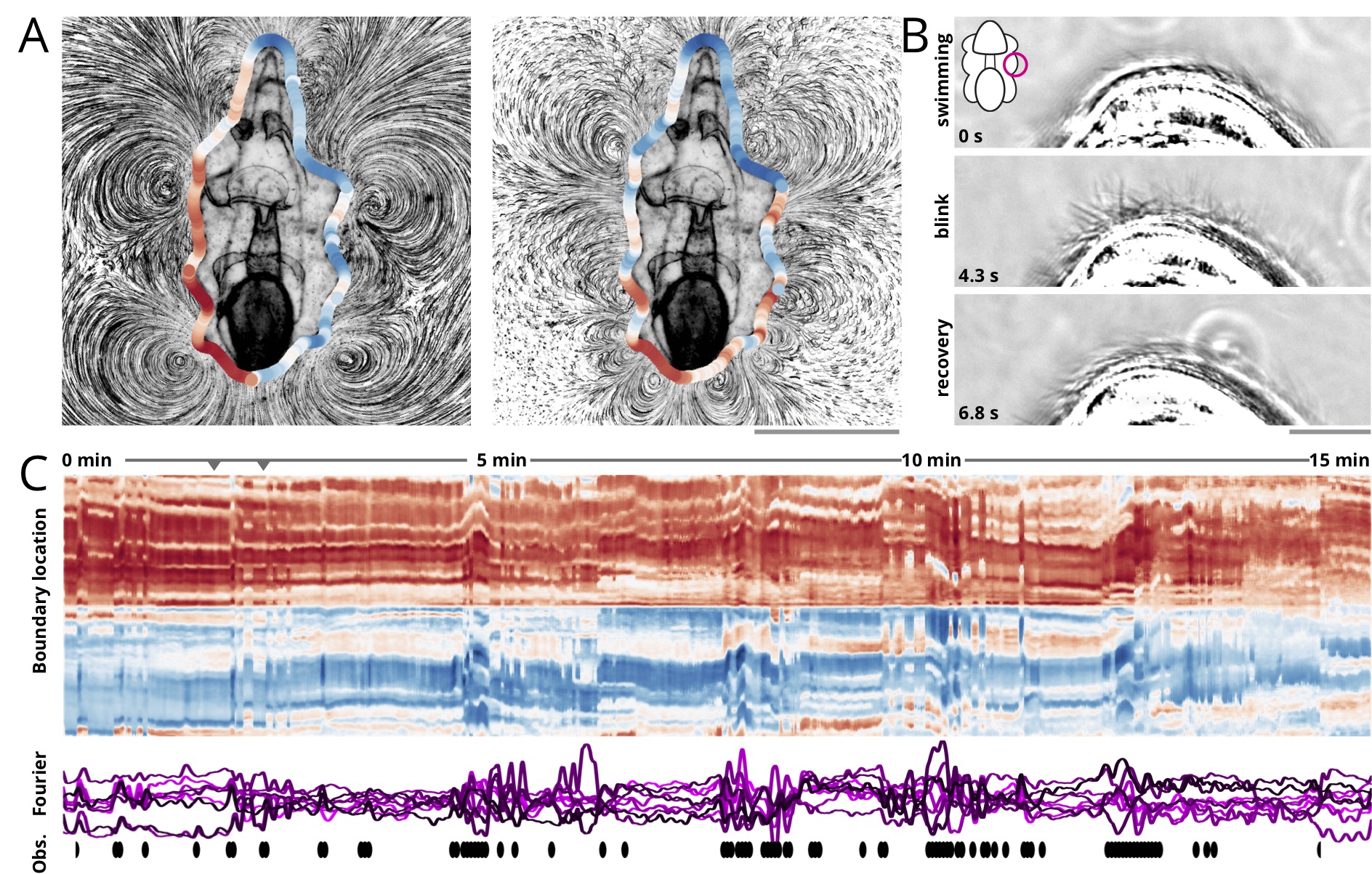}
\caption{
{\bf Quantitative behavioral decomposition reveals "blinks" in larval starfish swimming.} (A) Example flow fields produced by the larva beneath a slide during regular swimming (left) and a short-lived "blink" in the swimming stroke (right). The boundary conditions and larval body are superimposed. Pathline length $3$ s; scale bar: $300$ $\mu$m. (B) Near-field imaging of the ciliary band during a blink shows reversal and arrest of ciliary beating (see Supplementary Video 3). Scale bar: $50$ $\mu$m. (C) A kymograph of the boundary conditions (tangential velocity) imposed by the cilia on the surface of a larval starfish over a 15 minute video, colored by direction (clockwise in blue, counter-clockwise in red). Plotted below are the amplitudes of the first 10 Fourier sine coefficients of the boundary conditions, indicating the characteristic length scale of boundary condition fluctuations. The darkest trace corresponds to lowest-order coefficient $B_1$ and thus the swimming speed. Superimposed black ovals indicate when behavioral transitions are readily observed in the corresponding Supplementary Video 2. 
}
\label{behavior}
}
\end{figure}

We spawned and observed the larvae of {\it Patiria miniata}, a starfish with a freely-swimming larval stage that is canonical among marine invertebrates. Each larva has tens of thousands of beating cilia, which primarily localize in a band around the larva's periphery, an arrangement which facilitates body-wide coordination of beating by the larval nervous system. The larva can continuously adjust its swimming stroke through fine-grained control of its ciliary band, and in previous work we have shown that the larvae simultaneously swims and feeds on phytoplankton by modulating a complex array of vortices around its surface\cite{gilpin2017vortex}. Here we seek to analyze the dynamics of these vortices over long timescales in order to understand the larval behavior. We visualize vortices in 4-week-old larvae by seeding seawater with neutrally-buoyant, $0.75$ $\mu$m tracer particles, which we then film under dark field illumination for very long timescales ($15$ minutes) relative to known timescales of blinking ($< 1$ minute)\cite{gilpin2017flowtrace}. We observe continuous modulation of the swimming stroke, including during periodic blinking events in which the flow field globally shifts from a four-vortex "swimming" mode to a many-vortex "blinking" mode (Figure \ref{behavior}A; Supplementary Videos 1,2). During these blinks, individual cilia noticeably alter their beating dynamics (Figure \ref{behavior}B, Supplementary Video 3).

We next quantitatively analyze these behavior videos by extracting the local velocity field from each frame using particle image velocimetry (PIV). Because the quasi-two-dimensional flow field arises from the boundary conditions imposed by the ciliary band, in each frame we digitally segment the boundary of the larva and then computationally interpolate the PIV field onto it, resulting in a time series of flow velocities perpendicular ($\v v_\perp(t)$) and parallel ($\v v_\parallel(t)$) to the boundary. Across all datasets and boundary points, $\abs{\v v_\perp} \ll \abs{\v v_\parallel}$, suggesting preservation of the no-flux boundary condition despite the limited spatial resolution of PIV and potential out-of-plane components of the velocity field. The surface tangential velocity $\v v_\parallel(t)$ is centered and re-scaled by position relative to the larval anatomy, resulting in kymographs such as the one shown in Figure \ref{behavior}C that show transient reversal periods that produce the observed flow field variation. Because the larvae swim at low Reynold's number ($\text{Re} < 0.01$), these boundary condition kymographs contain all information necessary to reconstruct the entire flow field---and thus larval behavior---at a given time.

We next analyze behavioral motifs by computing the spatial Fourier sine coefficients $B_n$ of the boundary conditions via the identity $B_n = \int v_\parallel(\theta) sin(n \theta) d\theta$, where $\theta$ corresponds to the relative boundary position weighted by curvature (Figure \ref{behavior}C). The first term in this series, $B_1$, corresponds to a system-scale mode directly proportional to the larval swimming speed; the higher modes correspond to progressive smaller vortices in the near field \cite{gilpin2017dynamic}. Alternative orthogonal basis functions that better account for boundary geometry may be found empirically using standard manifold learning algorithms. We have observed that these functions primarily correspond to different length scales along the ciliary band, and thus strongly correlate with the more intuitive Fourier modes. We next classify temporal variation in the Fourier spectrum by applying principal component analysis to the time series of $B_n$. This reveals two characteristic combinations of Fourier coefficients, $\v B_{swim} \equiv (B_1^{swim}, B_2^{swim}, ...)$ and $\v B_{blink} \equiv (B_1^{blink}, B_2^{blink}, ...)$, associated with the two behavioral modes we observe in our videos. That the visually-complex flow fields observed at a given time readily decompose into characteristic superpositions of a finite number of surface Fourier coefficients illustrates the underlying low-dimensionality in the accessible behavioral space of starfish larvae\cite{berman2016predictability}, which likely arises from the strong physical constraints that act in the viscous environment. Using our mathematical model (described below), we compute stationary flow fields corresponding to $\v B_{swim}$ and $\v B_{blink}$, and find qualitative resemblance to the experimental flow fields in Figure \ref{behavior}A and other datasets\cite{gilpin2017dynamic}, suggesting that using two behavioral modes and a finite number of Fourier components captures the observed behavioral variation.

\begin{figure}
{
\centering
\includegraphics[width=.8\linewidth]{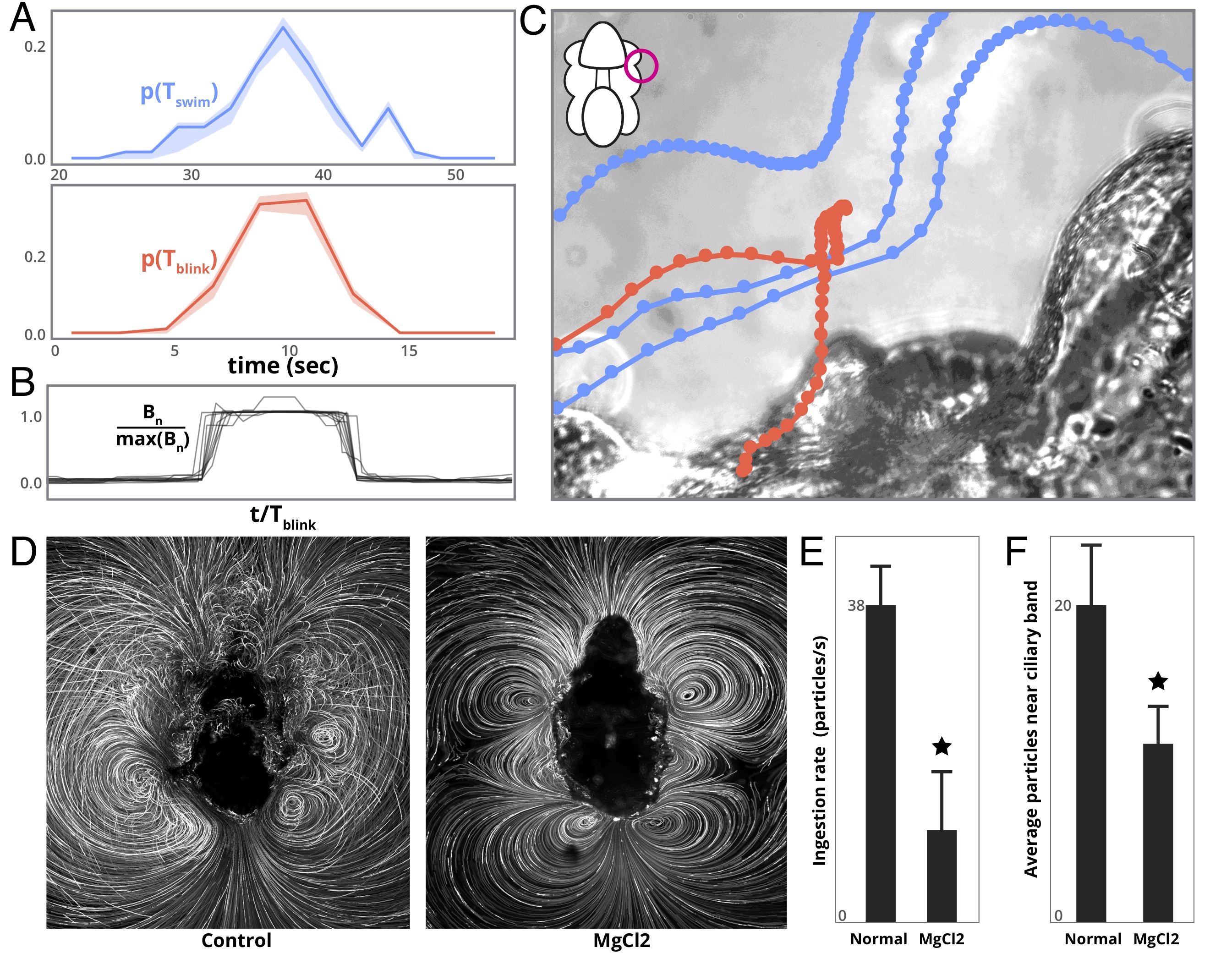}
\caption{
{\bf Rapid blinking transitions draw food particles towards larvae.} (A) Histograms of the distribution of swimming periods (upper blue curve) and blinking events (lower red curve) for $15$ organisms. Shaded ranges correspond to jackknifed distributions. (B) Single blinking events have a universal step shape when rescaled by their duration and amplitude. (C) A background-subtracted image shows tracks of particles passing near the body before (blue) and during (red) a blinking event.  (D) A $60$ s timelapse of a larva before (left) and after (right) treatment with the nervous-inhibitor MgCl$_2$. (E) The total rate of particles entering the mouth (the clearance rate) for untreated vs treated larvae. (F) The average number of particles within contact distance of the ciliary band for untreated vs treated larvae. Stars indicate Welch's t-test, $p < 0.001$, with $12$ organisms per condition.
}
\label{feed}
}
\end{figure}

The behavioral decomposition in Figure \ref{behavior}C shows that long stretches of steady swimming are punctuated by shorter blinks in which the swimming speed decreases and higher Fourier modes dominate. Averaging $15$-minute power spectra across $15$ organisms shows that blinking events have a surprisingly regular separation of $T_{swim} = 38 \pm 5$ seconds, and that they typically last $T_{blink} = 9 \pm 3$ seconds (Figure \ref{feed}A). These timescales are consistent across different $B_i$ and thus spatial scales, suggesting that they arise from organism-scale nervous coordination of the ciliary activity. Moreover, the shape of individual blinking events is extremely stereotypical when rescaled by duration ($t/T_{blink}$) and amplitude $B_i/\max(B_i)$ (Figure \ref{feed}B), further suggesting that these sharp behavioral transitions share a common underlying origin. Prior electrical measurements of invertebrate larvae have shown that similar-duration ciliary arrests are associated with a series of voltage spikes on the larval surface\cite{mackie1969electrical}; however, these spikes are suppressed when the ciliary band is isolated from the remainder of the larval nervous system\cite{mackie1976nervous}.

Because the ciliary band primarily exists to enable simultaneous locomotion and filter-feeding\cite{amemiya2015development}, we hypothesize that blinking plays a direct role in the larval feeding strategy. Normally, in order for larvae to intercept and ingest edible phytoplankton, the particles must pass close enough to the body to physically contact the ciliary band \cite{strathmann1971feeding,pernet2017larval}. We observe that during blinking events, some particles passing near the band abruptly change direction and travel orthogonally to their original trajectory and towards the ciliary band (Figure \ref{feed}C, Supplementary Videos 4,5). This suggests that regular blinking events periodically pull edible particles towards the larval surface, increasing the overall feeding rate. We quantify this effect by seeding the seawater around the organism with a mixture of strained algae culture water and $2$ $\mu$m yellow-green fluorescent beads. We expect the dissolved organic molecules from the strained algae water to mimic behavioral cues caused by natural algae, without introducing spurious imaging artifacts such as algae clumps or autofluorescence. We observe that these particles undergo erratic and uneven motions over $\sim$ $1$ minute timescales, and that many particles in the vicinity of the larvae are eventually captured and observed to pass into the animal's mouth region (Figure \ref{feed}D; Supplementary Video 6). However, surprisingly, when the larvae are first incubated for 30 minutes in seawater containing a low concentration of dissolved MgCl$_2$, these erratic behaviors disappear completely and the larval flow field becomes uniform (Figure \ref{feed}D; Supplementary Video 7; Supplementary Figure S1). MgCl$_2$ acts as a synaptic inhibitor in marine invertebrates\cite{mackie1969electrical}, leading to observable changes in the animal's swimming behavior\cite{strathmann1971feeding}; in fact, larvae exposed to MgCl$_2$ almost entirely lose their ability to reverse ciliary beating and change direction, resulting in pairs of animals colliding in the culture and becoming stuck together (Supplementary Video 8). 

Importantly, organisms exposed to MgCl$_2$ exhibit a substantial decrease in the overall rate at which they ingest particles, visible as decreased aggregation of particles in their stomachs (compare Supplementary Videos 9 and 10). By recording the integrated fluorescence intensity in oral region, we calculate that the "clearance" rate (particles ingested per second) decreases by nearly a factor of $4$ when behavioral modulation is inhibited by MgCl$_2$ (Figure \ref{feed}E). This suppression has previously been reported in invertebrate larvae\cite{strathmann1971feeding,hart1990manipulating}, however it has typically been attributed to individual cilia becoming unable to reverse and capture single passing particles\cite{mackie1969electrical}. To check whether this effect explains the observed feeding decrease, we use particle tracking and fluorescence to count separately the particles that pass close enough to the ciliary band to be captured, and we find that MgCl$_2$ nearly halves this quantity (Figure \ref{feed}F). This suggests that behavioral transitions affect the feeding rate through a hydrodynamic mechanism.

These experiments suggest that behavioral transitions play an important role in increasing feeding uptake, and they are consistent with previous experiments on echinoderm larvae feeding that counted the particles cleared by the ciliary band\cite{strathmann1971feeding,hart1990manipulating,pernet2017larval}. However, as in earlier experiments, immobilization of the larvae produces a mathematically-known recirculation effect in particles' pathlines over large length scales\cite{emlet1990flow,gilpin2017reply}. To leading order, these effects may be compensated numerically by subtracting a Stokeslet to account for the drag that a freely-swimming animal experiences\cite{gilpin2017vortex,gilpin2017dynamic}, although far from the animal's body additional considerations such as image forces and excess friction from the top boundary must be considered\cite{mathijssen2015tracer}. Importantly, the ciliary boundary conditions themselves are independent of confinement\cite{gilpin2017reply}, and so we use the experimental $\v v_\parallel(\theta,t)$ and $B_i(t)$ to formulate a model of low Reynolds number swimming in an unbounded domain.

\begin{figure}
{
\centering
\includegraphics[width=\linewidth]{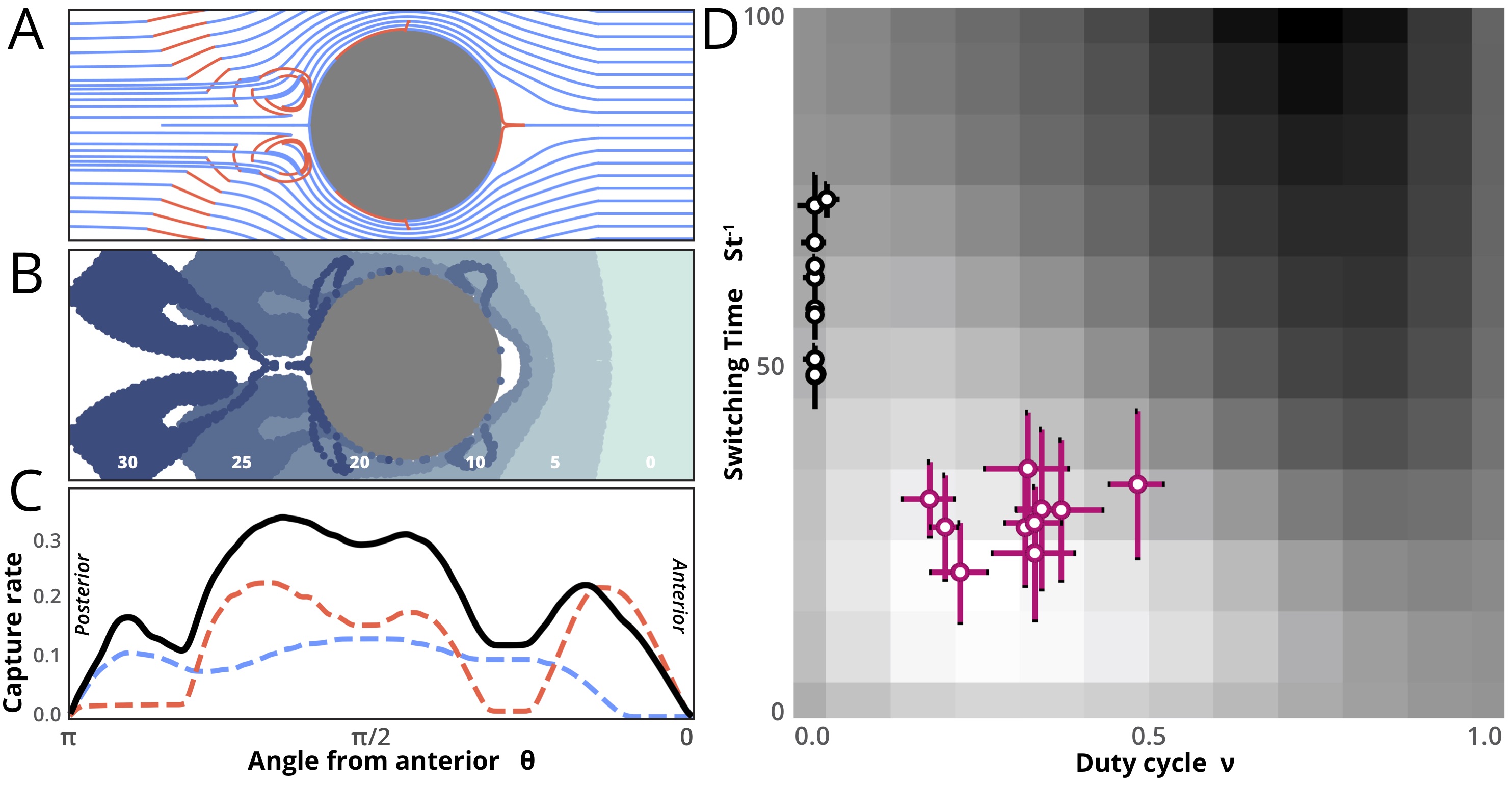}
\caption{
{\bf Enhanced feeding rate in a model blinking larva.} (A) Streamlines for a model swimmer that undergoes a single blink (red), with flow and time parameters taken from the average of the experimental datasets. (B) Simulated deformation of a cloud of particles encountered by the model swimmer during a blink, colored by time. (C) Relative rate at which particles are captured at each position along the swimmer perimeter (ciliary band). Comparison simulations with a swimmer locked in the blinking flow field (red) and swimming flow field (blue) are underlaid. (D) Simulated particle capture for transitions between the observed feeding and swimming modes, parametrized by the frequency of transitions $St^{-1}$ and the relative duration spent in each mode $\nu$. Color bar ranges from a capture rate $64\%$ below the mean across simulations (black) to $50\%$ greater than the mean (white). Overlaid points are experimental values of $(\nu, \text{St}^{-1})$ calculated from behavioral decomposition of $15$ minute videos of untreated (magenta) and MgCl$_2$-treated (black) organisms. Points correspond to distinct organisms, and error bars to observed ranges for each organism.
}
\label{model}
}
\end{figure}

For a circular swimmer with arbitrary slip boundary conditions, the flow field can be expressed using the time-dependent "squirmer" model\cite{blake1971self}, which describes the full flow field in terms of the surface Fourier coefficients $B_n(t)$,\begin{align}
v_r(r,\theta, t) &=  \dfrac12 B_1(t) \cos\theta \left(  \left(\dfrac{a}{r} \right)^2 - 1  \right) + \sum_{n=2}^{\infty} \dfrac{n}{2} B_n(t)\cos(n \theta)  \left(  \left(	\dfrac{a}{r} \right)^{n+1} -\left( \dfrac{a}{r} \right)^{n-1}	 \right) \notag\\
v_\theta(r, \theta, t) &= \dfrac12 B_1(t) \sin\theta \left(  \left(\dfrac{a}{r} \right)^2  + 1  \right) + \sum_{n=2}^{\infty} \dfrac{1}{2} B_n(t) \sin(n \theta)  \left(  n\left(\dfrac{a}{r} \right)^{n+1} -(n-2)\left( \dfrac{a}{r} \right)^{n-1}	 \right),
\label{squirmer}
\end{align}
where $a$ is the radius of the swimmer and the swimming speed is $v_{swim} = (1/2)B_1$. The behavioral variation we describe above simply consists of making all the Fourier coefficients functions of time, $B_i(t)$, in order to match the time series in Figure \ref{behavior}C. Here, we further simplify the behavioral dynamics by assuming that the larvae instantaneously shifts between the experimentally-computed parameter sets, $\v B_{swim}$ and  $\v B_{blink}$. We also assume that these transitions occur at evenly-spaced swimming intervals $T_{swim}$ and last for $T_{blink}$. Streamlines of oncoming particles corresponding to a single blinking event are shown in Figure \ref{model}A, indicating sharp turns and large particle displacements similar to those observed in the experimental videos.

We quantify feeding performance of the model "two mode" swimmer by numerically simulating its capture dynamics as it encounters a homogenous, horizontally-extant patch of particles. The blinking event distorts the algae patch and causes narrow filaments to form around the body, which trail the rest of the patch and thus linger near the body for an extended duration (Figure \ref{model}B; Supplementary Video 11). Using a longer algae path than the one shown in the panel, we can discard transient effects as the animal enters and exits the patch in order to focus solely on the effects of the blinking on feeding rate. From $N_p$ individual encountered particle trajectories, we calculate the time-resolved local density field, $\rho(\v r, t) \equiv \sum_i^{N_p} \delta(\v r - \v r_i)$. Because the starfish larvae feed by intercepting particles that contact their surface, we define an annular region of width $\delta$ near the body and record the average feeding rate $\bar f$ over the simulation,
\begin{equation}
\bar{f} \propto \lim_{t\rightarrow\infty}\left(\dfrac{1}{t}\int_0^{t}\!\!\! \int_0^{2 \pi}\!\!\! \int_{a}^{a+\delta}\!\!\!\!\!\! \rho(\v r, t') \;r dr d\theta dt'\right)
\label{intercept}
\end{equation}
We have assumed that the starfish larvae only capture a small fraction of particles that approach their surface\cite{strathmann1971feeding}, and so $\bar{f}$ scales with both the total number of oncoming particles that reach the organism's surface, and the net amount of time that they spend there. In the simulations, we also introduce a small "capture rate per unit time" corresponding to depletion of the local nutrient density, in order to prevent over-counting of particles that become trapped in eddies and linger near the surface---we choose this value based on ciliary band clearance rate estimates from the literature\cite{strathmann1971feeding,mackie1976nervous}, however our results do not strongly depend on the exact value of this parameter.

For the typical behavioral timescales used in the simulations, we find that $\bar{f}$ is roughly 2-3x larger than equivalent rates computed for swimmers without any behavioral variation ($T_{blink}=0$ or $T_{blink}=\infty$). These advantages arise jointly from an increased capture frequency at short timescales, and an increased utilization of the ciliary band due to refilling of depleted regions neighboring closed eddies. The latter effect is visible in the derivative under the integral of \eqref{intercept} with respect to $\theta$ (Figure \ref{model}C), which shows that the blinking feeder (black trace) captures more particles at more locations along the boundary than either stationary flow field (blue and red traces denote $\v B_{swim}$ and  $\v B_{blink}$). Analysis of the timing dynamics of a single blinking event (Supplementary Materials) further suggests that this advantage is transient, with the relative benefit of time-dependent swimming degrading after some time---illustrating the need for repeated, regular blinks.

We thus find that optimal feeding strategy blinks with sufficient frequency and duration that most incoming particles will experience a large enough deflection to draw them towards the surface along crossed streamlines. Blinking too frequently produces small excursions insufficient to cross streamlines, but blinking too rarely undermines the relative role of this transient contribution to the overall feeding rate. This motivates us to next compare how the overall feeding rate depends on the specific values of $T_{swim}$ and $T_{blink}$ by repeating the capture efficiency simulations for a range of values of these parameters (Figure \ref{model}D). We express the parameters in dimensionless form as the inverse Strouhal number, $ \text{St}^{-1} = (\bar{v}_{swim}/a)(T_{blink}+T_{swim})$ and the duty cycle $\nu = T_{blink}/(T_{blink}+T_{swim})$. The edges of parameter space, $\nu = 0$, $\nu = 1$, $\text{St}^{-1}=0$, $\text{St}^{-1}=\infty$ correspond to fixed swimming strategies without temporal variation. For this figure, simulations were performed with long enough times and a large enough algae patch that multiple blinking events could occur, leading to a resasonable estimate of the average feeding rates in a homogenous environment.

Consistent with intuition, the highest capture rates occur at intermediate values of $\text{St}^{-1}$ and $\nu$, where the swimmer spends sufficient time in each behavioral state that most oncoming particles experience large deflections from blinks.
Overlaid on the parameter space search are actual values of $(\nu,\text{St}^{-1})$ computed from the experimental videos; suggestively, these values appear near the theoretical maximum, while MgCl$_2$-treated organisms (which essentially spend no time blinking, $\nu=0$) are expected to have a much lower capture rate---explaining their reduced particle encounter rate in Figure \ref{feed}F.

\begin{figure}
{
\centering
\includegraphics[width=.5\linewidth]{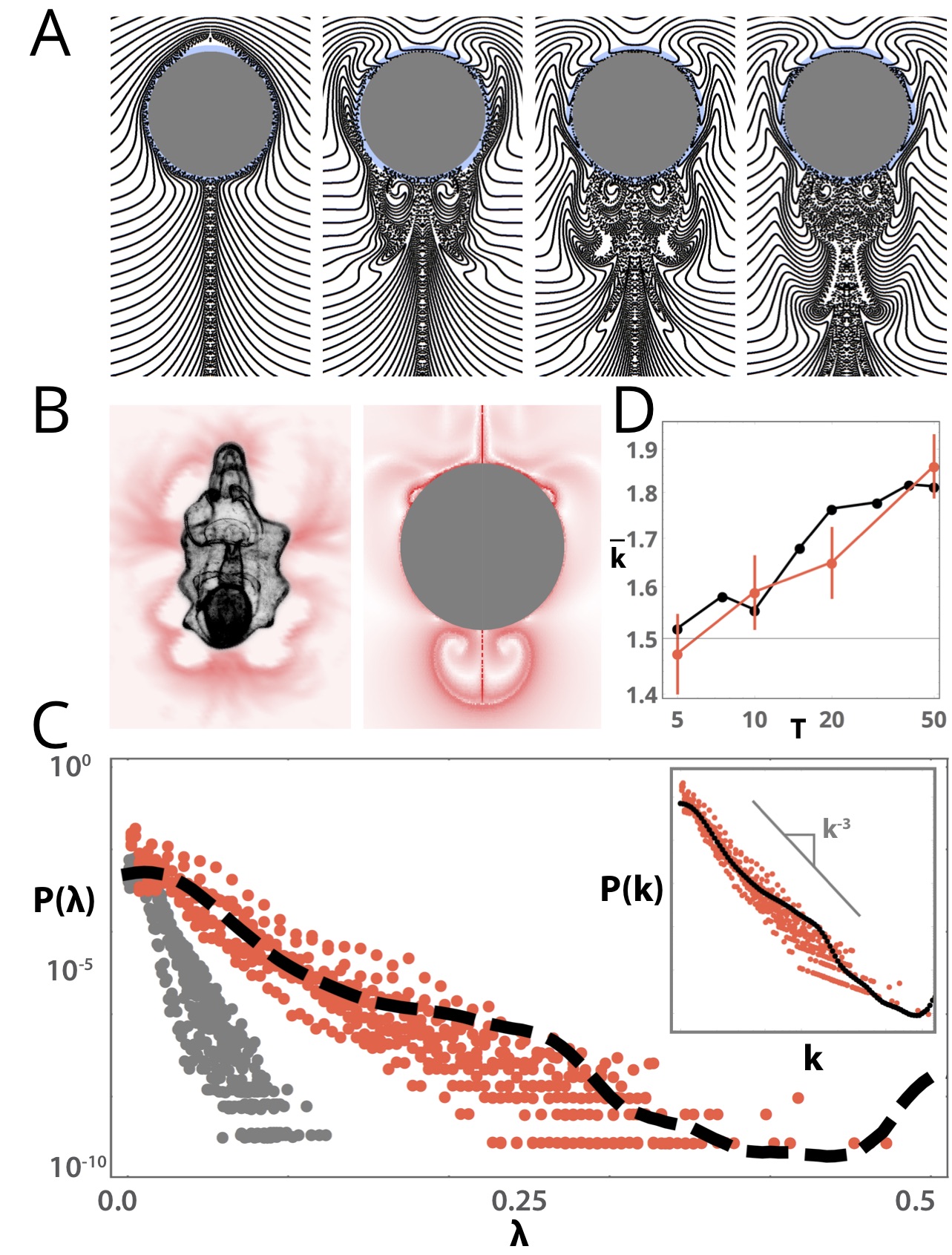}
\caption{
{\bf Chaotic mixing by a blinking feeder.} (A) Timelapse of the advection of uniform particles stripes encountered by a swimmer as it transitions from swimming behavior to a feeding behavior, illustrating the formation of a wake structure in the model (B) A plot of the largest finite-time Lyapunov exponent (FTLE) at each point in the experimental and modelled flow fields. (C) The distribution of FTLE values in the experimental (red) and theoretical (black) flow fields, with an equivalent analysis for MgCl$_2$ underlaid in gray. (Inset) The predicted power spectrum of a passive scalar advected by a flow field with this FTLE distribution. (D) The characteristic scale of the flow ($k$) as a function of timescale for the theoretical and experimental flow fields.
}
\label{chaos}
}
\end{figure}

Closer examination of the dynamics of trajectories during an optimal blinking cycle reveals that a complex far-field structure forms in the wake of the swimmer shortly after a blink (Figure \ref{chaos}A, Supplementary Video 12). This structure strongly resembles the "strange eigenmode" reported in several recent studies of chaotic mixing in open flows at low Reynolds numbers\cite{gouillart2011measures,lai2011transient,aref2017frontiers}. Model unsteady microswimmers have previously been observed to have wake structures consistent with chaotic stretching\cite{mueller2017fluid}; however, directly observing such mixing experimentally is challenging due to (1) rapid diffusion of passive tracers and (2) streamline distortion from confinement effects. Here, we analyze this effect using a combination of Lyapunov exponent analysis of our experimental velocity fields, and wake structure in our theoretical model.
 
Traditionally, chaos is characterized by a positive Lyapunov exponent, indicating that initially neighboring particles separate exponentially over time. From our experimental velocity fields, we measure this effect spatially using the largest finite time Lyapunov exponent (FTLE) \cite{peng2009transport},
\begin{equation}
\lambda(\v r) \equiv \dfrac{1}{\tau}\max\left[\text{eig}\left(			\log\left(	\d{\phi_{0}^{\tau}(\v r)}{\v r}	\right)		\right)\right]
\label{ftle}
\end{equation}
where the flow map $\phi_{0}^{\tau}$ transforms a tracer particle at initial position $\v r$ to its final position $\v r'$  after a time $\tau$. In the experimental flow fields, the FTLE distribution (Figure \ref{chaos}B) reveals the presence of filamentary structures that terminate on the larval surface. Such ridges typically correspond to isoclines of constant stretching, which serve as advective transport barriers outlining regions where particles preferentially linger relative to a steady flow\cite{haller2015lagrangian,gollubrothstein1999persistent}. These features are a necessary but not sufficient condition for chaotic advection\cite{peng2009transport,haller2015lagrangian}; however we find that they disappear when animals are treated with MgCl$_2$, implying that they require unsteadiness in the flow and thus chaotic effects. The FTLE ridges generally run orthogonal to steady vortex streamlines in the swimming-only mode (Figure \ref{behavior}), due to particles in these regions being deflected to a greater extent by blinks. For comparison, the FTLE field for the theoretical swimmer (Figure \ref{chaos}A) localizes more strongly to the near field and illustrates a filamentary structure in the swimmer's wake. This wake does not form in the experiments due to immobilization; however we expect its formation in freely-swimming animals due to the behavioral Fourier series in Figure \ref{behavior} that shows $\langle B_2(t)/B_1(t)\rangle_t>0$, consistent with a "puller" swimming stroke.\cite{mathijssen2015tracer}.

Over long timescales, the mixing properties of the starfish larval system may be analyzed through the probability distribution of FTLE values, $P(\lambda)$ (Figure \ref{chaos}C, red points). This distribution is substantially narrower for animals that have been treated with MgCl$_2$ (gray traces), indicating that behavioral variation directly facilitates mixing. Intriguingly, $P(\lambda)$ for the numerical model (dashed curve) almost exactly coincides with the experimental results without any fitting parameters, a surprising result given that qualitatively different streamlines arise in confined experiments compared to unbounded flow domains. This suggests that while immobilization affects the mixing geometry, the mixing dynamics arise from near-field effects and underlying intermittency in the dynamical timescales\cite{aref2017frontiers}. We next use recent results from mixing theory\cite{antonsen1996role} to transform $P(\lambda)$ into a leading-order estimate of the spatial power spectrum $P(k)$ for a continuous passive scalar advected by the flow\cite{fereday2004scalar,gollubrothstein1999persistent} (Figure \ref{chaos}D, inset). The nearly linear form of this decay suggests that the behavioral transitions produce a short-lived "forward cascade" in which portions of the nutrient field are deformed over progressively smaller length scales, corresponding to the filamentation regions visible in Figure \ref{model}B. Consistent with previous analyses of chaotic advection\cite{fereday2004scalar}, we find that this scalar cascade has similar mixing properties $P(k) \sim k^{-3}$ to two-dimensional turbulent flows at high Schmidt number\cite{bos2009inertial}. Repeating the FTLE analysis with longer integration times confirms that the mean wavenumber of mixing gradually increases as the local flow field stretches the expected tracer distribution (Figure \ref{chaos}D) \cite{gollubrothstein1999persistent}. Comparable low-Re mixing cascades and large tracer displacement distributions have previously been reported for bacterial turbulence\cite{leptos2009dynamics,wensink2012meso}, and here these effects suggest that local deflections observed during starfish blinks globally produce large-$k$ filaments of particles that linger near the surface and increase capture.

The appearance of chaotic advection in the starfish feeding behavior involves a characteristic combination of factors: streamline crossing, rapid time variation, and large average particle deflections near vortex regions\cite{aref2017frontiers}. For similar reasons, other filter-feeding systems (primarily sessile ciliates) have long been hypothesized to use chaotic effects to avoid depletion of the local nutrient field \cite{otto2001transport}; however, to our knowledge, ours is the first low-Reynolds number organism in which such behavior has been characterized experimentally. Because starfish larvae are freely-swimming, mixing in our system comprises "transient chaos"\cite{lai2011transient}, an analogous process to classical scattering wherein encountered particles only strongly interact with the chaotic saddle during their brief passage near the swimmer, before passing to the far field as part of a complex wake. Our results thus complement findings for larger-scale marine organisms in which inertially-driven mixing has been identified \cite{costello2008medusan,peng2009transport}. We further hypothesize that our observations have broader relevance to many other invertebrates with comparable larval forms and ciliary physiologies\cite{strathmann1971feeding,barnes2009invertebrates,pernet2017larval,riisgaard2010particle}, rapid behavioral changes\cite{conzelmann2011neuropeptides,mackie1969electrical}, and neural control systems\cite{mackie1976nervous,akira1989control}.

Mixing in the starfish larval system thus constitutes a direct example of how the timescale of an organisms' behavioral variation directly affects underlying physical forces---which may, in turn, affect survival through feedback on biological parameters like the overall feeding rate. Our findings contribute to a growing body of recent work suggesting tight coupling between environmental hydrodynamics and individual behavioral strategies in the ocean\cite{goldstein2016batchelor}; recent examples include diverse systems such as filtering by giant larvaceans\cite{katija2017new}, turbulence-mediated sand dollar larval settlement\cite{gaylord2013turbulent}, and active microbiome recruitment by bobtail squids\cite{nawroth2017motile}. Intriguingly, we observe little variation in behavioral dynamics in starved animals (Supplementary Figure S2); this is consistent with previous work on larval starfish suggesting that satiated or starved animals modulate feeding rate primarily via capture rates of single cilia along the band and the mouth, rather than organism-scale hydrodynamical changes\cite{strathmann1971feeding}. We thus suspect that blinking timescales have little behavioral plasticity, reminiscent of feeding rhythms observed in many basal invertebrates such as sponges and hydras\cite{passano1963primitive}. However, further experiments will be necessary to conclusively determine whether finer-scale statistical features in the behavioral time series respond to environmental variation; in particular, whether over longer timescales correlations emerge between behavioral states\cite{berman2016predictability} that could facilitate known life-history strategies like depth regulation or diel migration\cite{arkett1987neuronal,conzelmann2011neuropeptides}.

\subsection*{Acknowledgements}

We thank John O. Dabiri for helpful discussions and comments. We thank Guillermina Ramirez-San Juan, Scott Coyle, Arnold J. T. M. Mathijssen, and the Prakash lab for comments on the manuscript. This work was supported by a National Geographic Society Young Explorers Grant (W.G.) and the U. S. Department of Defense through the NDSEG Fellowship Program (W. G.), as well as an ARO MURI Grant W911NF-15-1-0358 and an NSF CAREER Award (to M.P.).

\newpage
\bibliography{blinking_cites} 
\bibliographystyle{naturemag}

\end{document}